\documentclass[showpacs,prl,twocolumn,superscriptaddress]{revtex4}
\usepackage{epsfig}
\usepackage{bm}
\begin{document}
\title{Orbital order in vanadium spinels}
\author{S. Di Matteo}
\affiliation{Laboratori Nazionali di Frascati INFN, via E. Fermi 40, C.P. 13,
I-00044 Frascati (Roma) Italy}
\affiliation{Dipartimento di Fisica, Universit\`a di Roma III, via della Vasca
  Navale 84, I-00146 Roma Italy}
\author{G. Jackeli}
\altaffiliation[Also at ] {E. Andronikashvili Institute of Physics,
Georgian Academy of Sciences, Tbilisi, Georgia.}
\affiliation{Institute for Theoretical Physics,
Ecole Polytechnique F\'ed\'erale de Lausanne, CH-1025, Lausanne,
Switzerland}
\author{N. B. Perkins}
\affiliation{Laboratori Nazionali di Frascati INFN, via E. Fermi 40, C.P. 13,
I-00044 Frascati (Roma) Italy}
\affiliation{Max-Planck-Institut f$\ddot{u}$r
Physik komplexer Systeme, N$\ddot{o}$thnitzer Str. 38 01187 Dresden, Germany.}
\affiliation{Bogoliubov Laboratory of Theoretical Physics, JINR, 141980, Dubna, Russia}

\date{\today}
\begin{abstract}
Motivated by recent theoretical and experimental controversy, we present a theoretical study 
to clarify the orbital symmetry of the ground state of vanadium spinel oxides AV$_2$O$_4$ (A=Zn, Mg, Cd).
The study is based on an effective Hamiltonian with spin-orbital superexchange interaction  and  a local 
spin-orbit coupling term.
We construct a classical phase-diagram and prove the complex orbital nature of the ground state.
Remarkably, with our new analysis we predict correctly also the coherent tetragonal flattening of oxygen octahedra. 
Finally, through analytical considerations as well as numerical ab-initio simulations, we propose how 
to detect the predicted complex orbital ordering through vanadium K edge resonant x-ray scattering.

\end{abstract}
\pacs{75.10.Jm, 75.30.Et }
\maketitle

Vanadium and titanium spinels, AB$_2$O$_4$ (B=Ti$^{3+}$, or V$^{3+}$), belong to a class of 
frustrated antiferromagnets where magnetic B-ions are characterized by orbital degeneracy due to 
partial occupancy of t$_{2g}$-orbitals (n$_{t_{2g}}$=1 for titanates and n$_{t_{2g}}$=2 for vanadates).
Recently  these spinels were thoroughly studied from both experimental \cite{onoda,lee,schmidt} and 
theoretical \cite{tsun,tcher,prl} 
points of view.
While the ground state of Ti-based spinels can be explained in terms of orbitally-driven superexchange
interactions on the frustrated pyrochlore lattice \cite{prl}, the situation seems not so fluid for vanadium spinels,
 as two conflicting
theoretical works appeared to explain their structural and magnetic properties \cite{tsun,tcher}.

In AV$_2$O$_4$, magnetically active V$^{3+}$-ions form a pyrochlore lattice
and are characterized by two $3d$ electrons in t$_{2g}$-orbitals, while A is a divalent ion like Cd$^{2+}$, 
or Zn$^{2+}$, or Mg$^{2+}$.
All compounds show qualitatively similar structural and magnetic behavior 
with a structural transition at a higher temperature $T_S$ and an antiferromagnetic (AFM) transition
at a slightly lower temperature $T_N$ \cite{10tcher}.
These findings have been interpreted by Tsunetsugu and Motome \cite{tsun}
as an interplay of $dd\sigma$ superexchange (SE) interaction and geometrical
frustration:  they showed  that ordering of orbitals can partially remove
magnetic frustration and explain the experimentally observed magnetic structure which is composed of AFM
 chains running in [110] and [1$\bar{1}$0] directions.
The ground state orbital ordering suggested in Ref.~[\onlinecite{tsun}] consists
of stacked $ ab$ planes with alternating  $d_{xz}$ and $d_{yz}$ vanadium hole orbitals (hereafter referred to as ROO).

On the other side Tchernyshyov \cite{tcher} pointed out that the ground state symmetry I4$_1$/a of 
ROO solution seems at odds with x-ray and neutron diffraction data, indicating a I4$_1$/amd space symmetry.
Thus, he proposed a purely ionic model where spin-orbit (SO) coupling plays the
major role and the V hole occupies (predetermining the sign of Jahn-Teller (JT) distortion) a complex linear combination  of $xz$ and
$yz$ orbitals: $(d_{xz}\pm i d_{yz})/\sqrt{2}$. (We shall refer to this orbital order as COO).

Actually, the correct space group of the system is still elusive. The tetragonal I4$_1$/amd space group 
was found in Ref.~[\onlinecite{onoda}], while
the authors of the neutron scattering experiment \cite{lee} supported ROO (thus, I4$_1$/a space group)
as more compatible with their findings. They found a clear signature of the one-dimensional character of 
spin-fluctuations in the temperature range $T_N<T<T_S$, indicating a weak coupling between AFM chains. 
Based on this observation, and claiming that interchain coupling is much stronger for COO compared to ROO, 
they interpreted the experimental results in favor of the latter.

The aim of the present paper is to shed some light on this issue.
To this end, we first link the two apparently unrelated pictures emerging from Refs.~[\onlinecite{tsun,tcher}]
and set them in an unique framework, by considering SE interaction and SO coupling  on an equal footing 
and constructing a classical ground state phase diagram.
We demonstrate that the SO coupling favors the states with unquenched orbital momentum
for any value of the coupling strength. We show that the only ground-state which is compatible with the 
experimentally observed magnetic structure is the COO-phase. This state is also characterized by a coherent 
tetragonal flattening of the oxygen octahedra along the same axis, as in real materials. 
Remarkably, this result is not pre-determined by any {\it ad hoc} choice of JT interaction 
as was done in Refs.~[\onlinecite{tsun,tcher}], but simply follows from our treatment of SE and SO interactions. 
We also evaluate the interchain spin couplings for both ROO and COO and find that they are comparable. 
Thus the experimental results of Ref.~[\onlinecite{lee}] cannot discriminate between the two types of orbital ordering. 
Finally, we suggest a possible key experiment to single out the proposed complex orbital states by means of 
magnetic resonant x-ray scattering (RXS).
To this aim, we performed a formal analysis and an ab-initio numerical simulation by means of the 
relativistic multiple scattering code in the FDMNES package \cite{joly,prbyves}.

{\it The effective Hamiltonian for AV$_2$O$_4$.} ---
The model Hamiltonian consists of two parts $H=H_{\rm SE}+H_{\rm SO}$. The first term represents 
the superexchange interaction between vanadium ions and the second term stands for the local spin-orbit coupling.
First we discuss $H_{\rm SE}$.
The insulating phase of AV$_2$O$_4$ is of Mott-Hubbard type and the SE interaction can be
described by the Kugel-Khomskii model \cite{kugel}.
We borrow the SE Hamiltonian for V$^{3+}$ ions, with three-fold degenerate $t_{2g}$ orbitals and spin S=1, 
from Ref.~[\onlinecite{v2o3}], to which we refer for technical aspects, and specialize it to our case.
Hund's exchange, $J_H \simeq 0.68$ eV, and Coulomb on-site repulsion (same orbital), $U_1\simeq 6.0$ eV, 
are taken from spectroscopy data \cite{V}. For future convenience, we introduce the small parameter 
$\eta =J_H/U_1\simeq 0.11$. Tight-binding fit to linear-augmented plane-wave calculations \cite{matt} 
shows that dominant overlap integrals are of $\sigma$-type with $t=3/4t_{dd\sigma}=-0.24$ eV \cite{nota1}.
For this reason, we consider only nearest-neighbor (NN) hopping integrals of
$dd\sigma$-kind on the pyrochlore lattice, as in Refs. [\onlinecite{tsun,prl}]. 
The  $dd\sigma$ overlap  in $\alpha\beta$ plane ($\alpha\beta$=$xz$, $yz$, $xy$)
connects only  the corresponding orbitals of the same $\alpha\beta$ type. 
Thus, orbital operators are Ising-like and their contribution can be expressed simply
 in terms of projectors $P_{i,\alpha\beta}$ onto the occupied orbital state $\alpha\beta$ 
at site $i$, which are equal to 1  when $\alpha\beta$ orbital is filled, or to 0 when it is empty.
We finally get: $H_{\rm SE}=\frac{1}{2} \sum_{ij} H_{ij}$, where

\begin{eqnarray}
H_{ij}=-{\big [}J_0 \vec S_i\cdot \vec S_{j}+J_1{\big ]} O_{ij}
-J_2{\big [}1-\vec S_i\cdot \vec S_{j}{\big ]}{\bar O}_{ij}
\label{spinorb2}
\end{eqnarray}

\noindent is the energy of bond $ij$. The orbital contributions $O_{ij}$ and ${\bar O}_{ij}$ 
along the bond $ij$ in $\alpha\beta$-plane are given by: $O_{ij}= P_{i,\alpha\beta}(1-P_{j,\alpha\beta})+
P_{j,\alpha\beta}(1-P_{i,\alpha\beta})$ and $\bar{O}_{ij}=P_{i,\alpha\beta}P_{j,\alpha\beta}$.
We defined: $J_0=\eta J/[1-3\eta]$, $J_1=J[1-\eta]/[1-3\eta]$, $J_2=J[1+\eta]/[1+2\eta]$, and 
the inequality $J_0<J_2<J_1$ 
holds. The overall energy scale is $J=t_{dd\sigma}^2/U_1\simeq 9.6$ meV.

The remaining local part of the effective Hamiltonian describes the relativistic spin-orbit coupling at each vanadium ion.
We adopt the usual representation of effective $\vec{L}'=1$
orbital angular momentum for $t_{2g}$-electrons \cite{ball}, with ${\hat {L}}'_z|xy\rangle=0$, and
${\hat {L}}'_z|xz\pm i yz\rangle/\sqrt{2}=\pm |xz \pm i yz\rangle/\sqrt{2}$. The true angular momentum $\vec{L}$ is related to 
the effective one by $\vec{L}\simeq-\vec{L}'$. With this notation $H_{\rm SO}$ can be written as:
 $H_{\rm SO}=-\lambda\sum_{i}\vec{L}'_i\cdot\vec{S}_i$, with $\lambda>0$.

\begin{figure}
\centerline{\epsfig{file=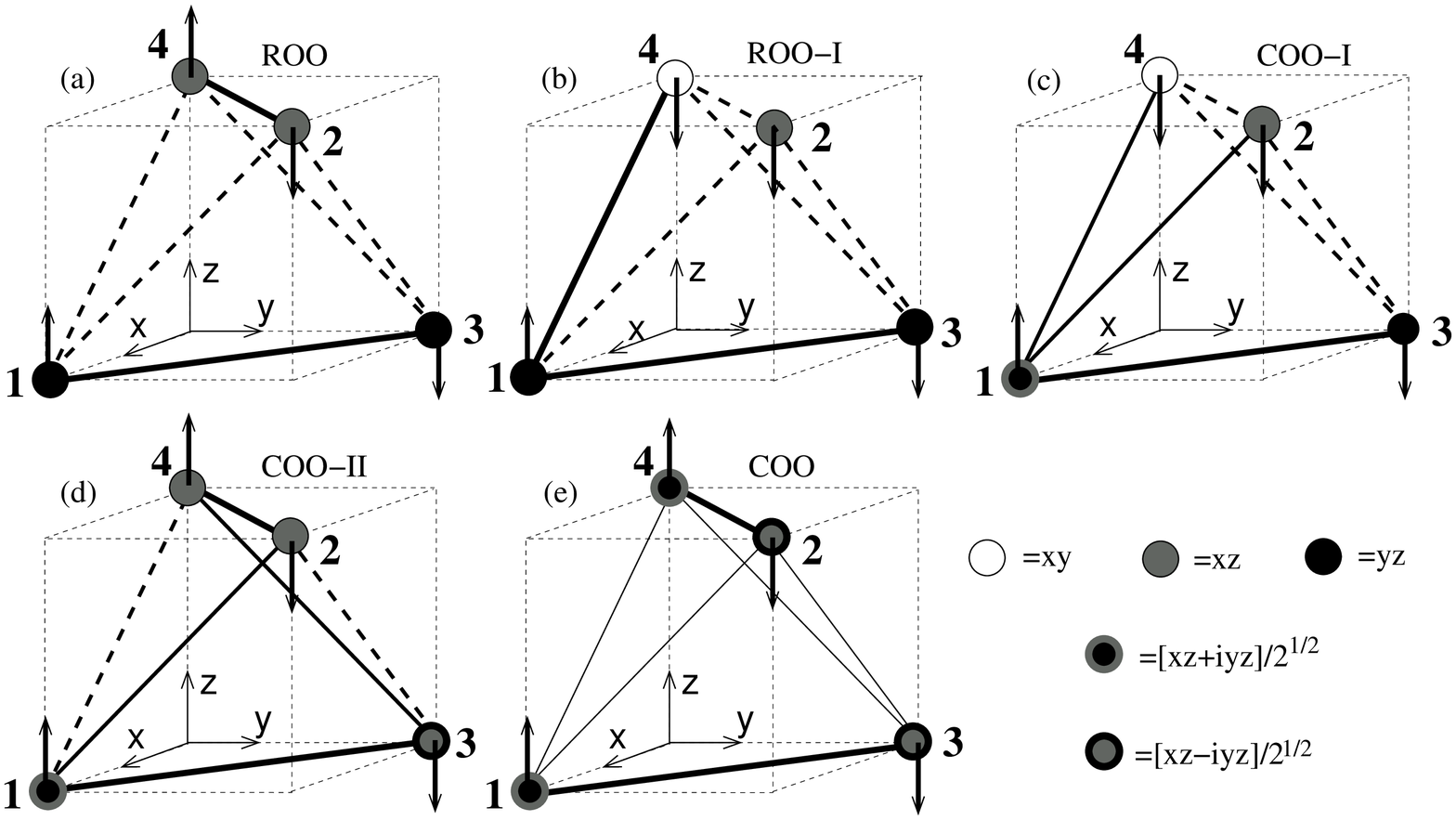,width=8.4cm}}
\caption{Orbital, spin  and bond arrangements on a tetrahedron. (a), (b), (c),
(d), and (e) respectively correspond  to
ROO, ROO-I, COO-I, COO-II, and  COO phases discussed in the text.
Hole orbitals are shown. The solid (dashed) lines stand for
AFM (FM) bonds. The thicker is the line the stronger is the coupling on the bond.}
\label{fig1}
\end{figure}

{\it Real orbital configuration}.--- We first consider the states when only real orbitals are occupied.
The orbital angular momentum is quenched and the 
local spin-orbit coupling is inactive.
Such orbital patterns are eigenstates of the orbital part of $H_{\rm SE}$.
This allows to classify all possible interacting bonds $ij$ in $\alpha\beta$ planes in two categories:
{\it i)} weakly ferromagnetic (FM) $b_1$-bond, when only one of the two sites has an $\alpha\beta$ electron (${O}_{ij}$ 
contribution); and {\it ii)} strongly AFM  $b_2$-bond with both $i$ and $j$ ions occupied by  $\alpha\beta$ electrons 
($\bar{O}_{ij}$ contribution). There are also non-interacting bonds when nor $i$ neither $j$ have a $\alpha\beta$
orbital occupied. As we are interested in magnetic configurations of the whole pyrochlore lattice, in the following 
we limit our considerations to classical spins.
The exchange coupling on $b_1$-bonds is given by $H_{b_1}=-J_{0}\vec S_i\cdot \vec S_{j}-J_1$, with
maximum FM energy gain: $E_{b_1}=J_0+J_1$.
For $\eta=0$ the bond is nonmagnetic and $E_{b_1}=J$.
The exchange coupling of $b_2$-bonds is given by $H_{b_2}=-J_2(1-\vec S_i\cdot \vec S_{j})$, and 
the maximum (AFM) energy gain is equal to $E_{b_2}=2J_{2}$. For future convenience we point out 
that $E_{b_2}=2E_{b_1}$ for $\eta=0$, while $E_{b_2}<2E_{b_1}$ for any finite $\eta$.

In order to identify the lowest energy state in the lattice, we introduce the
operators $N^{\rm T}_{b_1}=\sum_{\langle ij\rangle}^\prime  O_{ij}$ and
$N^{\rm T}_{b_2}=\sum_{\langle ij\rangle}^\prime{\bar O}_{ij}$ that count the number of interacting 
$b_1$ and $b_2$ bonds on a tetrahedron.
Using the identity for projectors $P_{i,xy}+P_{i,xz}+P_{i,yz}=2$
(as there are two electrons occupying different orbitals at each site), one obtains
$N^{\rm T}_{b_1}+2N^{\rm T}_{b_2}=8$.
Rewriting this  constraint in terms of energies of interacting $b_1$ and $b_2$
bonds, one finds that the energy gain per tetrahedron for $\eta=0$ is always equal to $JN^{\rm T}_{b_1}+2JN^{\rm T}_{b_2}=8J$ .
Therefore,  for $\eta=0$, all real orbital patterns are degenerate \cite{note}.
However, as with increasing $\eta$ the energy gain on $b_2$-bond decreases, while that on $b_1$ increases, the degeneracy is lifted. 
Thus, the lowest energy configuration is  the one with the maximum number of $b_1$ bonds per tetrahedron, 
which is $N^{\rm T}_{b_1}=4$ (recall that $N^{\rm T}_{b_1}+N^{\rm T}_{b_2}\leq$6-the total number of bonds on a tetrahedron).

\begin{figure}
\centerline{\epsfig{file=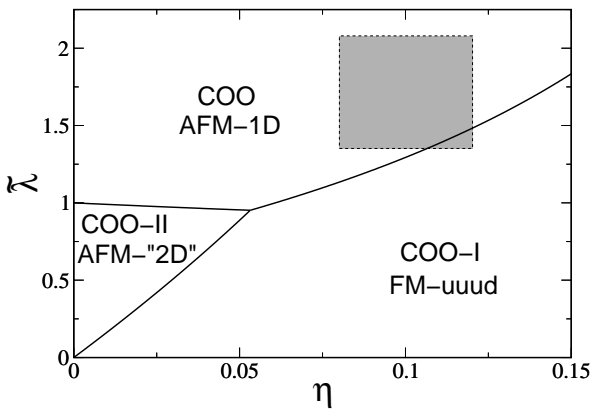,width=8cm}}
\caption{Classical phase diagram of $H_{\rm SE}+H_{\rm SO}$ in $(\tilde{\lambda},~\eta)$ plane.
The shaded rectangle denotes the domain of possible values of parameters.}
\label{fig2}
\end{figure}

There are two topologically different tetrahedral configurations
with $N^{\rm T}_{b_1}=4$ [see Fig. \ref{fig1}(a), \ref{fig1}(b)]. In the case of ROO (Fig. \ref{fig1}(a))
 there are
two strongly AFM $b_2$-bonds located on opposite edges and
coupled through weakly FM ($J^\prime=-J_0$), frustrated $b_1$-bond interactions.  
When all tetrahedra are of ROO-type, the average energy per site is $E_{\rm ROO}=-2J_1-2J_2$. 
The  magnetic structure can be viewed as a collection of AFM chains
running in [110] and [1$\bar{1}$0] directions, coupled through frustrated interchain interaction
$J^\prime$. At the classical level the chains  are decoupled from each other.

In the configuration ROO-I the two strong AFM bonds share a corner
and not all $b_1$-bonds are frustrated (Fig. \ref{fig1}(b)).
The spins
of each tetrahedron form a fully  collinear up-up-up-down ($uuud$)
state with a finite magnetic moment equal to 1/2 of saturation value.
The energy of this state $E_{\rm ROO-I}=-J_0-2J_1-2J_2$  
is the lowest among those with quenched angular momentum. 
Thus, ROO-I phase is the ground-state solution when $\lambda=0$. 
However, it is at odds with experiments in both magnetic structure and JT distortion. 
In fact, the latter should have the same local axis at all sites, while this 
is not the case for ROO-I phase.
Notice again that in Ref.~[\onlinecite{tsun}] ROO ground state was found because  JT coupling with only
tetragonal distortion in $z$-direction was pre-introduced. However, the inclusion of a local JT coupling with cubic symmetry
 will not affect the above results, as each state from the manifold would gain the same amount of JT energy by an elongation
of each VO$_6$ octahedron along the local tetragonal axis \cite{kugel}.

{\it Complex orbital configuration}.---  Now consider the orbital configuration shown in Fig. 1(c) (COO-I), obtained from ROO-I 
by replacing a $yz$ hole with $[xz+iyz]/\sqrt{2}$ at site 1. This site now carries angular momentum $L'_z=1$ and gains the SO
energy $\lambda$. The  spin couplings on bonds 1-2 and 1-4 are obtained by replacing orbital bond
operators $O_{12(4)}$ and ${\tilde O}_{12(4)}$ by their expectation values
$\langle O_{12(4)}\rangle=\langle{\tilde O}_{12(4)}\rangle=1/2$. This leads to the spin SE Hamiltonian
$H_{12(4)}=[-J_1-J_2+(J_2-J_0)\vec S_1\cdot \vec S_{2(4)}]/2$, with AFM coupling $[J_2-J_0]/2>0$.
Its ground-state energy is $E_{\rm COO-I}=-J_0-2J_1-2J_2-\lambda/4$, which is lower
than the lowest energy for real orbital patterns (ROO-I discussed above).
Therefore, for any finite value of $\lambda$ the states with quenched angular momentum are unstable with respect to those with unquenched $L_z'$. 
We can further construct the states with complex orbitals classifying them
by the number of sites with non-zero $L_z'$ per tetrahedron. The state with two complex sites (state COO-II, shown in Fig. 2(d)) 
has an average energy per site given by $E_{\rm COO-II}=-3J_2-J_1-\lambda/2$. It is lower than the energy of
COO-I for $\lambda>4[J_1-J_2+J_0]\sim8\eta$.  The magnetic order  for such an OO 
can be regraded as a quasi-two-dimensional  AFM pattern, formed by strong-bonds (see the path 1-2-4-3-1 in Fig. 1(d)), with weak 
FM interaction along diagonals (bonds 1-4 and 2-3).
The configuration with three complex sites has never the lowest energy. 
When all four sites carry non-zero $L'_z$ (state COO, shown in Fig. 1(e)), the ground state energy is $E_{COO}=-1/2[5J_2+2J_1]-\lambda$. 
The tetrahedron is characterized by two strong AFM $b_2$-bonds
(2-4 and 1-3) and the lattice magnetic structure is given by 1D AFM chains running in [110] and [1$\bar{1}$0] directions. 
The interchain coupling is frustrated and weakly AFM, with $J^\prime=1/4[J_2-2J_0]$, contrary to the ROO-phase whose $J^\prime=-J_0$ is FM. 
Its strength is practically the same for both COO and ROO. For example, for $\eta=0.12$, $|J^\prime|=0.14J$ for ROO and $|J^\prime|=0.16J$ for COO. 
Therefore, in both cases the magnetic network is quasi-one-dimensional
and the experimental results of Ref.~[\onlinecite{lee}] can not rule out one or the other, as we have already outlined.
Notice that the frustration of COO-phase is much weaker than ROO phases, as the spin direction is fixed along the $z$-axis by SO coupling and the 
degeneracy of the ground state is discrete as discussed in Ref.~[\onlinecite{tcher}].

In Fig. \ref{fig2} the ground-state phase diagram is shown in terms of the two dimensionless parameters $\tilde{\lambda}=\lambda/J$ and $\eta$. 
The shaded rectangle represents the domain of realistic values for parameters: $\eta=0.8-0.12$ \cite{tsun,V}, and $\lambda=13-20$ meV
(${\tilde\lambda}=1.35-2.08$) \cite{tcher,abrag}. 
Interestingly, if the system were close to the phase boundary,
the application of an external magnetic field might induce the
transition from COO, with zero magnetization, to COO-I, with 
a finite magnetic moment. 
The value of the field depends on how far from the phase boundary 
actual parameters locate the system.
For the parameters within the shaded rectangle in Fig. \ref{fig2} the maximal value of the critical field is estimated to be
$H_c\simeq 86$ T. The magnetic field experiment can be an interesting test of the phase diagram presented here.
 
To conclude the above consideration, we note that among ground states appearing in the phase diagram, 
only COO is compatible with the experimentally observed AFM structure.
This type of orbital ordering  also implies the coherent flattening of VO$_6$ 
octahedra along a unique axis leading to the experimentally observed tetragonal distortion. In what follows, we discuss a key experiment that we believe
can give unambiguous way to detect the suggested ground state orbital order.

{\it Experimental detection of COO-phase}.--- An effective way to unravel the question of the orbital symmetry, both below $T_N$ and between $T_N$ and $T_S$, 
is by means of V K edge RXS, where one
is sensitive to the on-site value of $L_z$, as the local transition amplitudes are added with a phase factor that compensates the vanishing effect due
 to the global symmetry\cite{nota4}.
The elastic scattering amplitude is $A(\vec{Q},\omega) = \Sigma _j e^{i\vec{Q} \cdot \vec{\rho_j}} f_j(\omega)$, where $\vec{Q}$ is the momentum transfer
 in the scattering process, $\hbar \omega$ the photon energy, $\rho_j$ the atomic positions in the unit cell, and the sum is over the 8 atoms
in the tetragonal cell. The atomic scattering factors (ASF), $f_j(\omega)$, is a second-order process in the electron-radiation interaction, 
given by (in the dipole approximation and atomic units) \cite{blume}:

\begin{equation}
f_j(\omega) = (\hbar \omega)^2 \sum _n \frac{\langle \Psi_0^{(j)}|{\hat {\epsilon}}_s^*\cdot \vec{r}|\Psi_n^{(j)} 
\rangle \langle \Psi_n^{(j)}|{\hat {\epsilon}}_i\cdot \vec{r}|\Psi_0^{(j)}\rangle}
{\hbar \omega - (E_n-E_0) -i\Gamma_n}
\label{asf}
\end{equation}

\noindent where $|\Psi_0^{(j)}\rangle$ ($|\Psi_n^{(j)}\rangle$) is the ground (excited) state, with the origin taken on 
the $j$-th scattering atom, and $E_0$ ($E_n$) is its energy; $\Gamma_n$ takes into account the finite lifetime of the 
excited states; $\vec{\epsilon}_{i(s)}$ is the incident (scattered) polarizations, and $\vec{r}$ is the coordinate of the 
electron in the reference frame of the resonant ion.
In COO-phase the space-group is I4$_1$/amd, and we can write the structure factor in terms
of the ASF of one single ion, through the symmetry operators $\hat{C}_{2z}$ and $\hat{C}_{4z}^{\pm}$ (two- and four-fold rotation around c-axis)
 and $\hat{T}$ (time-reversal): $S(\vec{Q}) =  {\big (}1+(-)^{h+k+l}\hat{T}{\big )}
{\big [} 1+(-)^{h+l}\hat{C}_{2z}+i^{h-k+l}\hat{C}_{4z}^++i^{h+k-l}\hat{C}_{4z}^- {\big ]} f_1$.
For the I4$_1$/a space group, induced by the reduced symmetry of the ROO-phase, we get:
$S(\vec{Q}) =  {\big (}1+(-)^{h+k+l}\hat{T}{\big )}{\big [} 1+(-)^{h+l}\hat{C}_{2z}+i^{-h+k+l}\hat{C}_{4z}^++
i^{-h-k-l}\hat{C}_{4z}^- {\big ]} f_1$. Therefore, for $h+k+l$=odd, both signals are proportional to $(1-{\hat T})$.
 As K edge magnetic signals are a signature of non-zero orbital angular momentum, 
we can use V K edge RXS to discriminate between ROO phase, where no signal is expected, and COO phase, where 
the signal is proportional to $L_z$ \cite{prbyves}.

\begin{figure}
\centerline{\epsfig{file=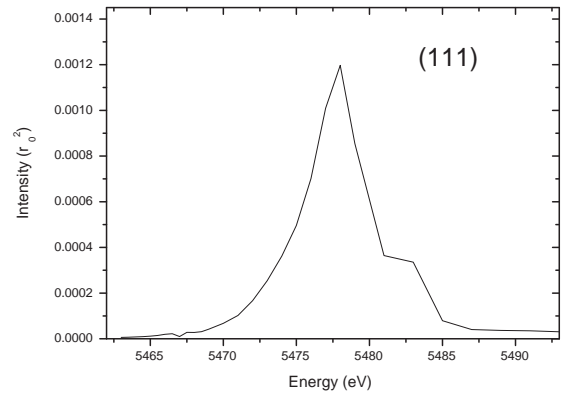,width=8.6cm}}
\vspace{-0.7cm}
\caption {Magnetic RXS at (111)$_{\sigma\pi}$ reflection for CdV$_2$O$_4$.}
\label{fig3}
\end{figure}
We have then performed a numerical simulation to estimate the order of magnitude of this signal for complex orbital occupancy with the relativistic multiple scattering code in the FDMNES package\cite{joly,prbyves}.
Results are shown in Fig. \ref{fig3} for the (111) reflection in CdV$_2$O$_4$ (structural data were taken from Ref. \cite{onoda}).
The incoming polarization is directed along one of $a$, $b$ axes, where the intensity is the highest. The signal is of the order of $10^{-3}r_0^2$
at dipolar energies ($r_0$ is the classical electron radius), which is well within the present capabilities of third generation x-ray sources: this is a key experiment
to choose the ground state in vanadate spinels. Probably a detectable signal could be obtained also through non-resonant magnetic x-ray scattering, thus allowing a quantitative measurement of $L_z$.



\begin{thebibliography}{99}
\bibitem{nota1}
The value $t_{dd\sigma}=-0.425$ eV given in Ref. [\onlinecite{matt}] has to 
be scaled by the ratio of the fifth power of the relative V-V distance, which is 3.07 {\AA} in, e.g., CdV$_2$O$_4$ [\onlinecite{onoda}] and 2.91 {\AA}
in LiV$_2$O$_4$, used in Ref. [\onlinecite{matt}].
\bibitem{note} We exclude the orbital patterns that generate frustrated spin exchange paths 
and thus higher energy states.
\bibitem{nota4}
Notice that a simple circular dichroism in absorption, as suggested in Ref. [\onlinecite{tsun}], 
is uneffective, as the average value of $L_z$ over the unit cell is zero.
\end{thebibliography}
\end{document}